\documentclass[lettersize,journal]{IEEEtran}

\usepackage{amsfonts, amscd, mathrsfs, amsopn, bbm, bbold, multicol, relsize}
\usepackage{mathrsfs}
\usepackage[psamsfonts]{amssymb}
\usepackage{mathbbol}
\usepackage[mathscr]{eucal}
\usepackage[cmex10]{amsmath}
\usepackage[amsmath,thmmarks]{ntheorem}
\usepackage{mathtools}
\usepackage{cite}
\usepackage{tikz}
\usepackage{graphicx}

\theoremstyle{plain}
\theoremheaderfont{\itshape\bfseries}
\theorembodyfont{\normalfont\itshape}
\theoremseparator{:}
\theoremsymbol{}

\theoremstyle{nonumberplain}

\theoremstyle{nonumberplain}
\theoremheaderfont{\itshape}
\theorembodyfont{\normalfont}
\theoremseparator{:}
\theoremsymbol{}

\theoremstyle{plain}
\theoremheaderfont{\itshape\bfseries}
\theorembodyfont{\normalfont}
\theoremseparator{:}

\theoremstyle{nonumberplain}
\theoremsymbol{\rule{1ex}{1ex}}

\newlength\fheight
\newlength\fwidth
\newcommand\restr[2]{{
  \left.\kern-\nulldelimiterspace 
  #1 
  \right|_{#2} 
  }}

\usepackage{amsmath,amsfonts}
\usepackage{algorithmic}
\usepackage{algorithm}
\usepackage{array}
\usepackage[caption=false,font=normalsize,labelfont=sf,textfont=sf]{subfig}
\usepackage{textcomp}
\usepackage{stfloats}
\usepackage{url}
\usepackage{verbatim}
\usepackage{graphicx}
\usepackage{cite}
\usepackage{lipsum}
\usepackage{adjustbox}
\usepackage{colortbl}
\usepackage{hhline}
\usepackage{float}
\usepackage{multicol}
\usepackage{hyperref}
\usepackage[section]{placeins}
\usepackage{setspace}
\definecolor{Gray}{gray}{0.88}
\definecolor{Grayy}{gray}{0.77}
\definecolor{Grayyy}{gray}{0.82}
\usepackage{multirow}

\def\BibTeX{{\rm B\kern-.05em{\sc i\kern-.025em b}\kern-.08em
		T\kern-.1667em\lower.7ex\hbox{E}\kern-.125emX}}

\newcounter{CagkanCounter}

\newcounter{RonCounter}

\begin{document}

\title{Dataset of Pathloss and ToA Radio Maps With Localization Application}


 


\author{
\IEEEauthorblockN{%
\c{C}a\u{g}kan~Yapar\IEEEauthorrefmark{1},
Ron~Levie\IEEEauthorrefmark{2},
Gitta~Kutyniok\IEEEauthorrefmark{3}\IEEEauthorrefmark{4}\IEEEauthorrefmark{5},
and Giuseppe~Caire\IEEEauthorrefmark{1}
}
\\
\IEEEauthorblockA{\IEEEauthorrefmark{1}Technical University of Berlin}%
\\
\IEEEauthorblockA{\IEEEauthorrefmark{2}Technion – Israel Institute of Technology}%
\\
\IEEEauthorblockA{\IEEEauthorrefmark{3}Ludwig Maximilian University of Munich}
\\
\IEEEauthorblockA{\IEEEauthorrefmark{4}University of Tromsø}

\IEEEauthorblockA{\IEEEauthorrefmark{5}Munich Center for Machine Learning (MCML)}
}



\maketitle

\begin{abstract}
In this article, we present a collection of radio map datasets in dense urban setting, which we generated and made publicly available.  The datasets include simulated pathloss/received signal strength (RSS) and time of arrival (ToA) radio maps over a large collection of realistic dense urban setting in real city maps. The two main applications of the presented dataset are 1) learning methods that predict the pathloss from input city maps (namely, deep learning-based simulations), and, 2) wireless localization. The fact that the RSS and ToA maps are computed by the same simulations over the same city maps  allows for a fair comparison of the RSS and ToA-based localization methods.
\end{abstract}

\begin{IEEEkeywords}
data set, radio map, pathloss, received signal strength (RSS), time of arrival (ToA), localization, machine learning, deep learning.
\end{IEEEkeywords}

\section{Introduction}\label{sec:Introduction}




\IEEEPARstart{R}{adio maps} are representations of quantities of interest for wireless communications applications, in a fine spatial grid. One traditional quantity of interest is the so-called pathloss.

The \emph{pathloss} (or \emph{large-scale fading coefficient}), quantifies the loss of wireless signal strength between a transmitter (Tx) and receiver (Rx) due to large scale effects. The signal strength attenuation can be caused by many factors, such as free-space propagation loss, penetration, reflection and diffraction losses by obstacles like buildings and cars in the environment. In dB scale pathloss amounts to $\textup{P}_{\textup{L}} = (\textup{P}_{\textup{Rx}})_{\rm dB}-(\textup{P}_{\textup{Tx}})_{\rm dB}$, where $\textup{P}_{\textup{Tx}}$ and $\textup{P}_{\textup{Rx}}$ denote the transmitted and received locally averaged (over multipath, small-scale phenomena) power (also called Received Signal Strength - RSS) at the Tx and Rx locations, respectively. Throughout this paper, we also use the term \emph{pathgain}, interchangeably with pathloss.
 
Many applications in wireless communication explicitly rely on the knowledge of the pathloss function, such as device-to-device (D2D) link scheduling, or user-cell site association. For example, in the latter, the goal is to assign a set of wireless devices to a set of cellular base stations, and in order to decide which device to assign to which station, it is important to know the radio map.
 
 Some other use cases of pathloss radio maps are fingerprint-based localization methods where the fingerprint (FP) is the signal strength (RSS) from different ``anchor'' infrastructure nodes (e.g., base stations or access points) \cite{LocUNetTWC}, 
 physical-layer security, power control in multi-cell massive MIMO systems, user pairing in MIMO-NOMA systems, precoding in multi-cell large scale antenna systems, path planning, and activity detection (see e.g. the references in \cite{RadioUNetTWC}).

In the following, we present datasets of simulated radio maps based not only on pathloss but also on time of arrival (ToA), at each point in a fine spatial grid, in large dataset of city maps.
To the best of our knowledge, there is currently no such publicly-available dataset.
The fact that we extract the signal strength and ToA from the same simulation in each of the environments is a  key-point,  which enables using our dataset for fair comparisons of  wireless communication methods that are based on the RSS and those that are based on ToA. We hence believe that our dataset will be useful for the research community in this area.

The high accuracy of the software we used for our simulations (the ray-tracing software WinProp from Altair \cite{WinPropFEKO})  was demonstrated by field measurements in many cities such as Helsinki, Munich, Nancy, Stuttgart, and Hong Kong, (see \cite{IRTMANET,DPM} and the references therein). Moreover, such simulation methods are frequently used by e.g. cell operators, proving their efficiency.

Using such computer simulations allowed us to generate high volumes of data, through which many studies are made possible. Moreover, applications like wireless localization, which has been an important motivation for our endeavours, benefit from the high resolution of the radio maps, as the grid size implies a minimum error in the accuracy of localization. Our presented datasets with their 1 m resolution are suitable for studies of high accuracy localization methods, whereas constructing such radio maps through measurement campaigns would be extremely difficult.



\begin{figure}[!t]
    \centering
    
        \includegraphics[width=.45\textwidth]{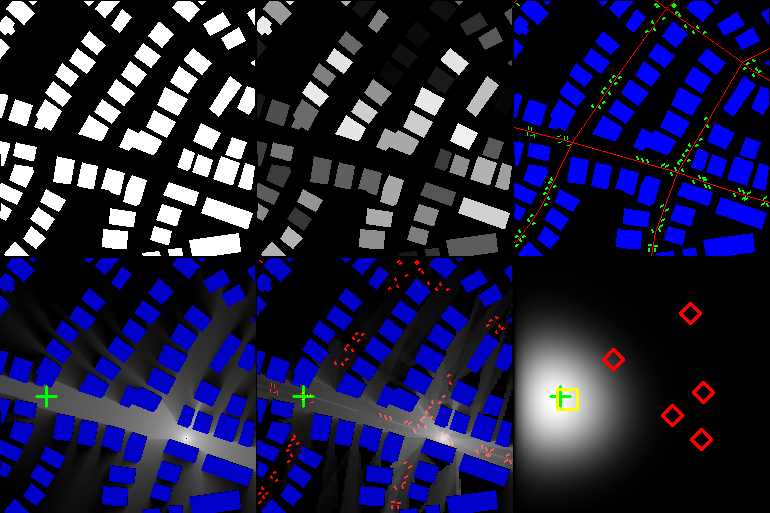}
        \caption{A visual overview of the presented datasets and their application in localization. Shown are: City maps w/o and w/ height encoding, roads and cars, pathloss radio map prediction and simulation under different settings, and the result of a localization experiment with LocUNet \cite{LocUNetTWC}.}
        \label{fig:allTogether}
   
\end{figure}

In the following, we explain the fundamentals on which all the presented datasets \footnote{The datasets are available at \url{https://ieee-dataport.org/documents/dataset-pathloss-and-toa-radio-maps-localization-application}\\Cagkan Yapar, Ron Levie, Gitta Kutyniok, Giuseppe Caire, October 7, 2022, "Dataset of Pathloss and ToA Radio Maps with Localization Application", IEEE Dataport, doi: \url{https://dx.doi.org/10.21227/0gtx-6v30}.} are based, followed by their individual descriptions and use cases. A visual overview of the datasets and their use in localization is shown in Fig.. \ref{fig:allTogether}.

\section{General Setting}\label{sec:Common}



We employ a collection of propagation models, with various complexities, given by \emph{WinProp}\cite{WinPropFEKO}, on a dataset of urban environments. 
We consider 701 city maps of size $256 \times 256$ meters, which were fetched from \emph{OpenStreetMap} \cite{OpenStreetMap} from the cities Ankara, Berlin, Glasgow, Ljubljana, London, and Tel Aviv.
In each map, 80 transmitter locations are considered, amounting to a total of 56080 simulations for each propagation model.  We divide the simulations to two types: \emph{ground level}, where the 80 Tx are located at ground level (1.5 m) for each map, and \emph{rooftop}, where the 80 Tx are located on the rooftops of the buildings. The former simulations correspond to the D2D setting, where the simulation is essentially 2D, and the latter simulations correspond to the base-station setting, where 3D considerations are essential for the simulation. 



We used two types of radio map simulation methods -- \emph{Dominant Path Model (DPM)} \cite{DPM} and \emph{Intelligent Ray Tracing (IRT)} \cite{IRT}. In IRT simulations, the maximum number of interactions of the rays with the environment is a parameter that effects the complexity/accuracy of the simulations. We used IRT with 2 and 4 interactions, which we call \emph{IRT2} and \emph{IRT4}, respectively. All buildings (and cars) are assumed to have the same generic material property, for simplicity.

Another parameter of the IRT simulations that affects the fineness of the reflection patterns is the length of the segments/tiles of the objects (buildings and cars in our datasets).

All simulations were saved with a resolution of 1 m per pixel, as \texttt{.png} images. Furthermore, the data describing the simulation settings, i.e., the transmitter locations, city maps, and cars (when applicable) are provided as images, together with their corresponding coordinates/shapes (as polygons) in \texttt{.json} files. The roads are saved as images and as polygonal lines.
\looseness=-1

%
   The pathloss values in the dataset were truncated  below a minimum pathloss value, and the range between this minimum pathloss value and the maximum pathloss value over the whole dataset was scaled to gray levels between 0 and 255, to save the pathloss radio map simulations as images.  
Notice the importance of putting emphasis on high pathgains, since the applications that use the pathloss values generally would favor communication links with higher pathgains. Hence, high pathgain values should be represented by gray levels close to 255 (maximum).

Also, note that one should only consider the points  where the RSS $(\textup{P}_{\textup{Rx}})_{\rm dB} = \textup{P}_{\textup{L}}+ (\textup{P}_{\textup{Tx}})_{\rm dB}$  lies above the noise floor $(\mathcal{N})_{\rm dB}$, such that the transmitted signal can be reliably detected. In other words, pathloss values which lead to RSS below the noise floor, should be omitted. In the Appendix \ref{sec:AppTrunc}, we briefly showcase how we found our truncation values (which we called \emph{pathloss threshold} and \emph{analytic pathloss threshold}), which might be beneficial for users of the dataset.
%
    The parameters and the found pathloss thresholds  are summarized in Table \ref{table:parameters}.

    \begin{table}[!t]	
	\renewcommand{\arraystretch}{1}
	\centering
        \caption{Parameters of the datasets}
	\scalebox{0.8}{
	\begin{tabular}{cc}
		\rowcolor{Gray}		\bfseries Parameter &  \bfseries Value\\
  \rowcolor{Grayy}\multicolumn{2}{c}{\bfseries \quad \quad Common Parameters in the 2D and 3D Datasets}\\
	    Map size &  $256^2$ pixels\\
		Pixel length &  1 meter\\
		  Rx  height &  1.5 meters\\
		
		Noise power spectral density &  -174 dBm/Hz\\
		Transmit power &  23 dBm\\
            Antenna type &  Isotropic\\
  \rowcolor{Grayy}\multicolumn{2}{c}{\bfseries \quad \quad Parameters of the 2D Datasets}\\
  Building height &  25 meters\\
  Tx  height &  1.5 meters\\
Center carrier frequency & 5.9 GHz\\
		Channel bandwidth &  10 MHz\\
  Noise figure & 0 dB\\
            Min PL in the simulations  & -186 dB\\
            Max PL in the simulations/dataset & -47 dB\\
		PL threshold &  -127 dB\\
            PL range &  80 dB\\
         PL threshold (analytic, dataset) &  -147 dB\\
            Simulation type &  DPM / IRT\\
            Max no. of interactions in IRT &  2 / 4\\
		
            Tile length in IRT simulations & 100 / 10 meters\\
             Simulations with cars in the map & No / Yes\\
   \rowcolor{Grayy}\multicolumn{2}{c}{\bfseries \quad \quad Parameters of the 3D Dataset}\\
   Height range of the buildings &  6.6-19.8 meters\\
            Tx height &  3 m above the rooftop \\
   Center carrier frequency & 3.5 GHz\\
            Channel bandwidth &  20 MHz\\
            Noise figure & 20 dB\\
            Min PL in the simulations  & -162 dB\\
            Max PL in the simulations/dataset & -75 dB\\
            
		PL threshold &  -104 dB\\
            PL range &  29 dB\\
            PL threshold (analytic, dataset) &  -111 dB\\
            
            Simulation type &  IRT\\
            Max no. of interactions in IRT &  2\\
            Tile length in IRT simulations & 10 meters\\
            Simulations with cars in the map & No 
	\end{tabular}
	}
	 \label{table:parameters}
\end{table}
\section{2D Datasets}\label{sec:2D}

In this section, we present datasets in the 2D setting, i.e., Tx  deployed in the street level, 1.5 m above ground, having the same height as the $256\times 256$ receiver pixels.  

We first introduce the \emph{RadioMapSeer Dataset}, the first pathloss radio map dataset that we generated and used in our RadioUNet work \cite{RadioUNetTWC}.  
Then, we present two additional  IRT datasets, which were generated to provide radio maps with more fine-grained reflection patterns, in the same city maps and Tx locations as in \emph{RadioMapSeer}. 
Last, we present two datasets (pathloss and ToA radio maps) designed specifically for   wireless localization application. 

For consistency and fair comparisons, in all 2D datasets, the same city and car maps, and the same Tx locations were used. Tx were restricted to be positioned within the $150 \times 150$ area in the center of the $256 \times 256$ city map.


\subsection{2D Radio Map Versions with Cars}

For each simulation models (i.e., DPM and IRT under various complexities) in 2D scenarios, we provide a pathloss radio map dataset version with the presence of cars in the environment. The specifications of these simulations are summarized as follows.

\begin{itemize}
    \item For each of the 701 city map, 100 cars  of size $2 \times 5 \times 1.5$ (width $\times$ length $\times$ height) are added (if the roads of the corresponding maps are of sufficient length to fit them, otherwise the maximum number that can fit), with adopting a random generation procedure near and along or perpendicular to roads. Specifically, we uniformly randomly picked road segments and randomly placed 
cars either parallel or perpendicular to the chosen segments. We note that length of the segments are
different and as a result shorter segments have more cars per segment length in general. This
provides in fact a more realistic placement of the cars, since shorter segments are mostly due
to curves in roads and more traffic congestion in curvy roads is expected.
     \item We calculated the effects of cars by running DPM simulations with the presence of cars, subtracting such results from DPM simulation results without cars, and by scaling this difference by $0.5$. We arrived at this modeling decision as a result of a discussion with an expert.
    \item The found effect of cars are then added to the radio maps simulated by the mentioned methods to generate their versions with cars.
\end{itemize}

\vspace{-1.5mm}

\subsection{The RadioMapSeer Dataset}
 In this dataset, the simulation parameters were chosen as follows.

\begin{itemize}
    \item DPM in the generic simulation setting, run for the 701 city maps with 80 different Tx locations (a total of 56080 simulations).
    \item  IRT with 2 (either a diffraction or reflection) and 4 maximum interactions of rays , where the tiling size was set to 100 m. We consider 80 and 2 Tx per map for IRT2  and IRT4 respectively, i.e., a total of 56080/1402  IRT2/IRT4 simulations.
\end{itemize}

To represent scenarios where the city map is inaccurate, additional versions of the considered 701 city maps are provided, where in each map of the original city map, one to four randomly chosen buildings are removed. Six experiments of random removals for each city are provided.

The main application of \emph{RadioMapSeer Dataset} studied in detail in our RadioUNet \cite{RadioUNetTWC} was RSS radio map estimation, e.g. achieving high accuracy radio maps predictions, but in much shorter time than a ray-tracing software. 
The presented method can take two versions of inputs. In the first case, only the city map and the Tx location image are given as inputs, similar to a simulation setting of a ray-tracing software. In the second version,  samples from the ground truth radio map are used as additional input features. This version allows one to adapt to the real propagation phenomena, when e.g. the available simulation method (used for the initial supervision) or the environment map is not accurate enough. Note that, in both cases, RadioUNet can yield very accurate results, in previously unseen city environment and Tx deployment scenarios.\looseness=-1

One interesting observation is  worth mentioning here: When studying the scenarios with missing buildings in the city maps, RadioUNet demonstrated an ability to infill their pixels, when RSS values from the radio map were sparsely sampled to compensate for the mismatched knowledge about such buildings. Here, the detection of the missing buildings itself can be of interest, exemplifying an unexplored potential use case of the presented dataset. Further details and other example applications of this dataset can be found in \cite{RadioUNetTWC}.


\subsection{Additional Higher Accuracy 2D Pathloss Radio Map Datasets}
	
Two more pathloss radio map datasets with finer reflection patterns were later generated, which were obtained with IRT simulations with more precision and higher computational complexity. Such finer patterns with sharp pathloss transitions can be of interest for some sensitive applications. 
In these pathloss radio map datasets, the tile length of the elements in the map   (buildings, and cars if applicable) is reduced from 100 m (which resulted in having only one tile for the most of the building walls) to 10 m, promoting the finer reflection phenomena in the simulations. 
%

 The first dataset is based on IRT2 simulations, with multiple interactions, i.e., a total of 2 maximum interactions for rays (recall the IRT2 setting in RadioMapSeer allowed for either a reflection or diffraction), whereas the second dataset was obtained with IRT4 simulations.


\subsection{Datasets Specialized for Localization}

We have also generated two new datasets to allow for comparisons among RSS (pathloss) and ToA ranging-based localization algorithms in realistic dense urban settings, which we called \emph{RadioLocSeer Dataset} and \emph{RadioToASeer Dataset}.

In dense urban areas, localization methods that rely on distance (range) estimations (between Tx and the Rx to be located) experience drastic performance deterioration, because of the unavoidable errors in the range estimations in an urban environment. Such errors occur due to the interactions of the rays with the obstacles present in the environment, such as the buildings, cars, pedestrians, incurring signal strength loses and delays, which results in very inaccurate distance estimations. Hence, radio map (fingerprint)-based methods are preferred over the methods which explicitly make use of such (mismatched) distance estimations. 

The presented \emph{RadioLocSeer Dataset} was designed to study RSS (pathloss) radio map-based localization methods.

\subsubsection{RadioLocSeer Dataset}\label{subseq:Dataset_RadioLocSeer}

An important motivation to generate this dataset was that localization methods do not have access to the ground truth radio maps, and instead make use of fast pathloss radio map estimation methods, such as the deep learning method \emph{RadioUNet} \cite{RadioUNetTWC} (Please see \cite{LocUNetTWC} for details). Hence, in the presented localization datasets we provide estimated radio maps generated by RadioUNet. Here, RadioUnet was supervised on ground truth DPM radio maps over 602 city maps, and then evaluated on 99 test-set  city maps. For each ``test-set'' city map and Tx location we provide simulated pathloss radio maps  under all the different simulations we explained for the 2D setting. This dataset contains a total of 7920 such radio map estimations per simulation setting, in 99 city maps and 80 Tx locations per map.  Being specialized for the localization task, the pathloss radio maps in this dataset are provided as truncated at the found \emph{pathloss threshold}, i.e., at -127 dB (cf. Table \ref{table:parameters}), instead of the \emph{analytic pathloss threshold} (cf. Appendix \ref{sec:AppTrunc}). DPM radio map estimations by RadioUNet \cite{RadioUNetTWC} in this subset (which was unseen during its training) of maps (amounting to a total of $99 \times 80 = 7920$ estimations) are also included in the dataset. 

 	\begin{table}[!t]
	\renewcommand{\arraystretch}{1}
	\centering
 \caption{\small Comparison of the performance of the pathloss radio map-based LocUNet \cite{LocUNetTWC} with ToA ranging-based methods. Mean absolute error accuracies of the compared algorithms under different additional noise and number of Tx in the map.}
    \scalebox{0.7}{
	\begin{tabular}{c|c|c|c|c}
	\hline
		\rowcolor{Gray}  {\cellcolor{Grayyy} \bfseries  no Tx:}&$\mathbf{5}$&$\mathbf{3}$&$\mathbf{5}$&$\mathbf{3}$\\
		\hline	
		\rowcolor{Gray}  {\cellcolor{Grayyy} \bfseries  Standard dev. of noise:}&\multicolumn{2}{c|}{ $\sigma = 0$} & \multicolumn{2}{c}{$\sigma = 20$}\\
		\hline
		POCS \cite{POCSgholami2011wireless} & $37.75$  & $46.15$ &  $41.27$ & $48.72$ \\
		\hline		
		SDP  \cite{SDP}& $\mathbf{6.81}$  & $\mathbf{13.95}$   & $24.88$ & $41.02$ \\
		\hline
		Robust SDP \cite{SDPR}, $b=20$ & $9.98$ & $17.10$ &  $27.76$ & $40.37$ \\
		\hline
		Robust SDP \cite{SDPR}, $b=0.7$ & $7.04$ & $15.38$ &  $28.42$ & $41.74$ \\
		\hline
		Bisection-based robust method \cite{BisecRob}, $b=20$ & $9.16$ & $15.87$ & $\mathbf{23.30}$ & $\mathbf{38.14}$\\
		\hline
		Bisection-based robust method \cite{BisecRob}, $b=0.7$ & $9.49$ & $14.95$ & $24.09$ & $40.75$ \\
		\hline\hline
		LocUNet \cite{LocUNetTWC} Nominal $\&$ Robustness  &  $\mathbf{4.80}$&$\mathbf{10.70}$  &  $\mathbf{13.14}$& $\mathbf{19.06}$ \\
	\end{tabular}
	}
	 \label{table:ToAAll}
\end{table}


	\subsubsection{RadioToASeer Dataset}\label{subseq:Dataset_RadioToASeer}
For the ranging based methods, ToA information is considered to be more preferable over RSS based-ranging, under the assumption of having access to hardware and protocols that allow for precise ToA measurements. This dataset allows studying such methods in the urban setting. 

This dataset was generated based on DPM simulations, of the same settings as in \emph{RadioLocSeer}, which provides ToA information of the dominant paths, to allow for fair and consistent comparisons between RSS and ToA ranging-based methods in realistic urban scenarios. 

 In the Appendix \ref{sec:AppToAOpt}, we explain how the ToA value of a dominant ray path (evaluated from the DPM simulation) constitutes a (quasi) lower bound on the true ToA. This ultimately means that evaluating the accuracy of ToA localization methods on \emph{RadioToASeer} gives optimistic (upper bound) estimates of the actual accuracy of an ToA ranging-based localization method. 

 Examples of the usage of these datasets can be found in \cite{LocUNetTWC}, where state-of-the-art RSS and ToA-ranging based methods were compared, to the best of our knowledge, for the first time in the literature. Moreover, in these works we presented a RSS radio map and deep learning-based method called \emph{LocUNet}, which essentially uses the pathloss radio map estimations (e.g. from RadioUNet for fast results) and the RSS measurements of the Rx from the Tx. We report in Table \ref{table:ToAAll} the numerical comparisons between the LocUNet and the state-of-the-art ToA ranging-based methods (cf. \cite{LocUNetTWC} for the details), demonstrating LocUNet's superior performance over the ToA ranging-based methods in various settings. 

 Based on these datasets, we organized a Data Competition at MLSP 2023 \cite{MLSP}.

   \setlength{\medskipamount}{5.2pt plus 0.1pt minus 0.1pt}

\begin{figure}[!t]
    \centering
    
        \includegraphics[width=.48\textwidth]{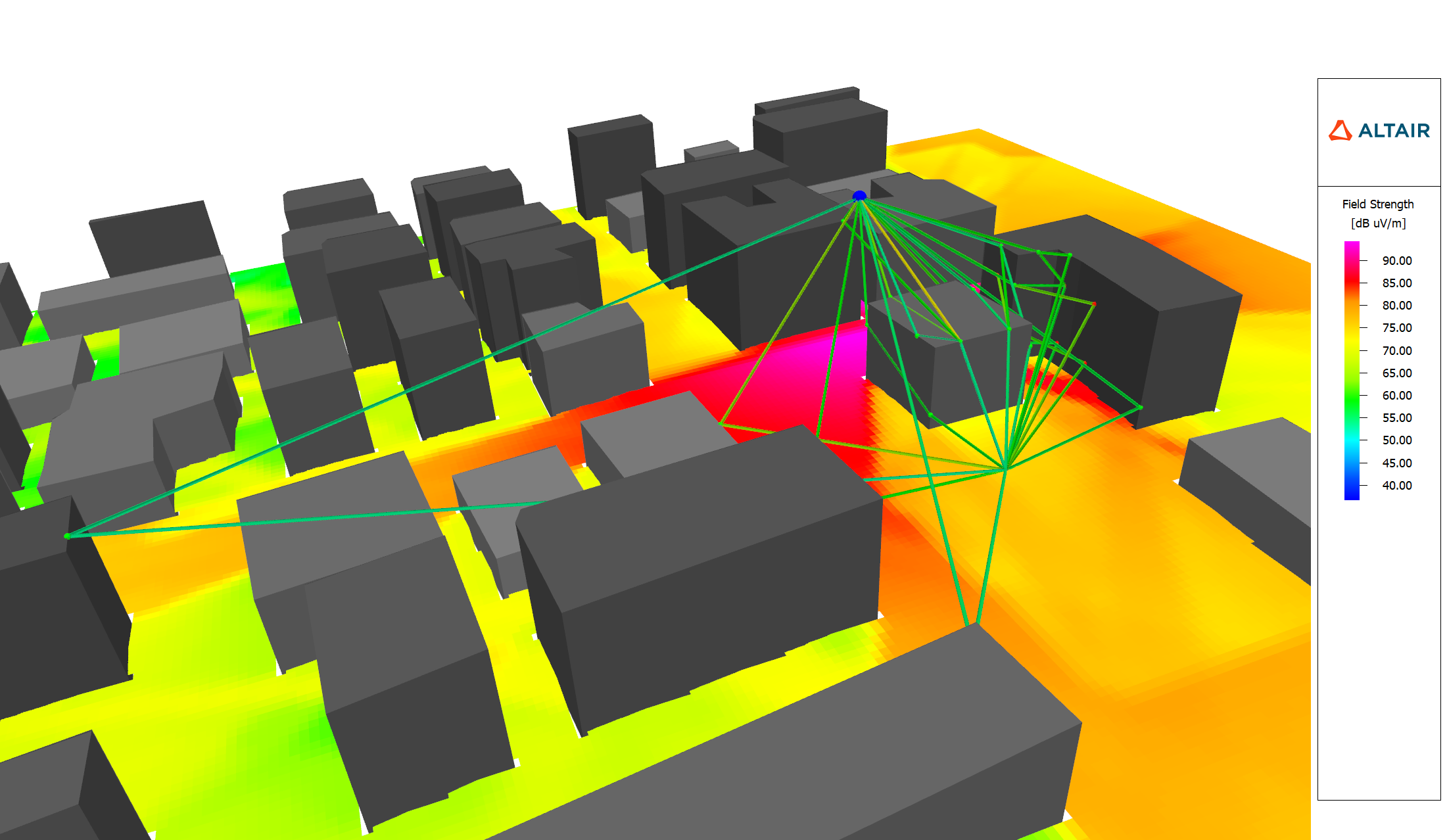}
        \caption{A simulated map from the presented 3D pathloss radio map dataset. The rays arriving at a chosen pixel are shown. Tx mounted on the rooftop of a high building}
        \label{fig:3D_Rays}
   
\end{figure}

\section{The RadioMap3DSeer Dataset}\label{sec:3D}

Last, we extend our simulations to the 3D setting, where varying building heights and Tx deployment on rooftops are considered. This dataset could find use in studying e.g. cellular scenarios, where the transmitters are placed on the rooftop. Notice that over the rooftop propagation gives rise to richer and more complicated pathloss patterns than in the 2D case (see Fig. \ref{fig:3D_Rays} for a simulated radio map example, which is also contained in our dataset after the appropriate post-processing). Thus, to study or  design deep learning methods which perform well in the 3D setting, 3D radio map datasets are required. The presented 3D dataset has the following specifications:

\begin{itemize}
\item The simulations are conducted in the setting of IRT with multiple ray interactions (max. 2), with 10 m tiling length of the building elements.
\item The same (as in the previous 2D cases) 701 $256 \times 256$ city maps are used. Each building in a city map is assigned a height that lies between 2 to 6 stories, where a story is taken as 3.3 m. This range of 13.2 m (from the minimum of 6.6 meters to the maximum of 19.8 m) is divided into 255 equal length levels and building heights are found by picking one of these levels uniformly. This data is provided in two image sets, one as black and white (BW) images of the pixels occupied by buildings, and one with their encoded height as gray levels. As in the previous datasets, the corresponding polygons (2.5D) in \texttt{.json} format are provided.  
    \item Transmitters are generated on the buildings that have a height of at least 5 stories (16.5 m). The transmitters are placed close to the edges to reflect the realistic deployment. The transmitter height from the rooftop is set to 3 m. We have restricted the Tx to be positioned within the $150 \times 150$ area in the center of the $256 \times 256$ city map if possible, and considered a larger area of $230 \times 230$ for the city maps when this was not possible, due to lacking buildings (above which the Tx could be deployed) in the center.

\end{itemize}

\begin{figure}[!t]
    \centering
    
        \includegraphics[width=.48\textwidth]{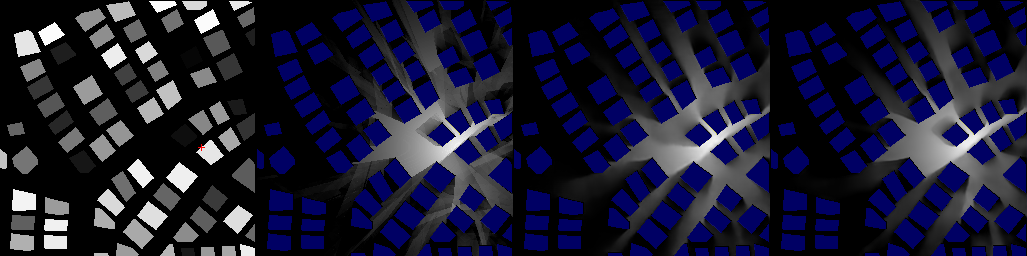}
        \caption{An example application of the 3D dataset. Shown are the height-encoded city map and the Tx (red plus sign), 3D radio map from the dataset, and the radio map predictions by the naive and the 3D adapted RadioUNet \cite{RadioUNetTWC} variants.}
        \label{fig:3DTogether}
  \vspace{-4mm} 
\end{figure}

In Fig. \ref{fig:3DTogether}, we show an example use of the presented dataset, where we trained RadioUNet with two versions of input: In the first one (naive), only the black-and-white 2D images of the buildings (city map) and Tx  were used as input features, whereas in the second case, the height information of the buildings were used as additional input features, through an appropriate usage of the provided height-encoded city images. In particular, we have decomposed the 3D building maps into equal length 2D horizontal \emph{slices}, such that each slice corresponds to an interval $\left[\textup{H}_{\textup{slice,min}},  \textup{H}_{\textup{slice,max}}\right]$ in the vertical direction, and the height of the buildings are re-scaled in this interval, such that heights below and above this range acquire 0 and 1, respectively, while the intermediate values lie within $\left(0,  1\right)$. We observed that 12 equal length slices together with the  BW (2D) city map and the Grid Anchor maps introduced in \cite{transRadio} yielded a good accuracy with RMSE 0.87 dB (cf. Fig. \ref{fig:3DTogether}, right), outperforming the naive approach, which had resulted in an RMSE of 1.26 dB.

The example in Fig. \ref{fig:3DTogether} demonstrates the capability of the 3D-adapted method to learn the effect of different building heights, on the pathloss radio maps, as witnessed by the different shadow lengths appearing behind the buildings with different heights, complying with the simulations by the ray-tracing software and real propagation phenomena. However, with the naive approach, such differences are lost.

Based on this dataset, we organized a Signal Processing Grand Challenge \footnote{\url{https://RadioMapChallenge.GitHub.io/} (open for late submissions).} at ICASSP 2023 \cite{Challenge,ChallengeLong}.


\section*{Acknowledgments}
We thank Ibrahim Rashdan for a fruitful discussion on the impact of cars on the pathloss function. The work presented in this paper was partially funded by the DFG Grant DFG SPP 1798 “Compressed Sensing in Information Processing” through Project Massive MIMO-II, and by the German Ministry for Education and Research as BIFOLD (ref. 01IS18037A). 

\vspace{-4mm}
\begin{appendices}

\label{appendix}

\section{Determining the Radio Map Truncation Values}\label{sec:AppTrunc}
In the following, we provide a demonstration of how we found the pathloss truncation values (cf. Sec. III-B in \cite{RadioUNetTWC} for more details).
First note that one should consider the points  where the received signal power $(\textup{P}_{\textup{Rx}})_{\rm dB} = \textup{P}_{\textup{L}}+ (\textup{P}_{\textup{Tx}})_{\rm dB}$  lies above the noise floor $(\mathcal{N})_{\rm dB} $, i.e. the points where $(\textup{P}_{\textup{Rx}})_{\rm dB} \geq (\mathcal{N})_{\rm dB}$ holds, where $(\mathcal{N})_{\rm dB} =  10\log_{10} W N_0 + \textup{NF}$ is the noise floor in dB, with NF being the noise figure. Solving this for pathloss $\textup{P}_{\textup{L}}$ we get the \emph{pathloss threshold} $\textup{P}_{\textup{L,thr}}$ as
 $\textup{P}_{\textup{L}}\geq \textup{P}_{\textup{L,thr}} = -(\textup{P}_{\textup{Tx}})_{\rm dB} + (\mathcal{N})_{\rm dB}.$ Even though any signal below the noise floor cannot be detected in reality, some applications might benefit from having pathloss simulation values that lie below the pathloss threshold (e.g. the coverage classification application presented in \cite{RadioUNetTWC}). Hence, the radio maps in the presented dataset are truncated at a lower threshold $\textup{P}_{\textup{L,trnc}} < \textup{P}_{\textup{L,thr}}$, where we chose $\textup{P}_{\textup{L,trnc}}$ such that the difference between the maximum pathloss $M_1$ in the dataset and $\textup{P}_{\textup{L,thr}}$ is approximately four times greater than the difference between $\textup{P}_{\textup{L,thr}}$ and $\textup{P}_{\textup{L,trnc}}$, i.e., $M_1-\textup{P}_{\textup{L,thr}} = 4(\textup{P}_{\textup{L,thr}}-\textup{P}_{\textup{L,trnc}})$. We dub $\textup{P}_{\textup{L,trnc}}$ the \emph{analytic pathloss threshold}. 
    Considering all the above mentioned points, the pathloss values in the radio maps were calculated by $f=\max\{\frac{\textup{P}_{\textup{L}}-\textup{P}_{\textup{L,trnc}}}{M_1-\textup{P}_{\textup{L,trnc}}},0\}$, with $M_1$ denoting the maximal pathloss in all simulated radio maps. Hence, $f=0$ represents anything below the analytic pathloss threshold, and $f=1$ represents the maximal pathgain at the transmitter. We note that there might be other considerations in the link budget calculation, such as a minimum required SNR level, which could be similarly incorporated in the above calculations. Following the above ideas, users of the presented datasets who are interested in settings with more stringent requirements, such as higher SNRs, can truncate the radio maps at higher threshold values.

\section{On the (Quasi) Optimality of the ToA Dataset}\label{sec:AppToAOpt}

In this section, we argue that evaluating ToA-based localization algorithms using \emph{RadioToASeer} yields upper bounds for their performances in real deployment. Please see \cite{LocUNetTWC} for a more detailed discussion.

First, we note that the dominant path is the shortest free space path connecting the Tx and the Rx.

\begin{figure}[!t]
		\vspace{-2mm}
		\centering
		\includegraphics[width=0.75\linewidth]{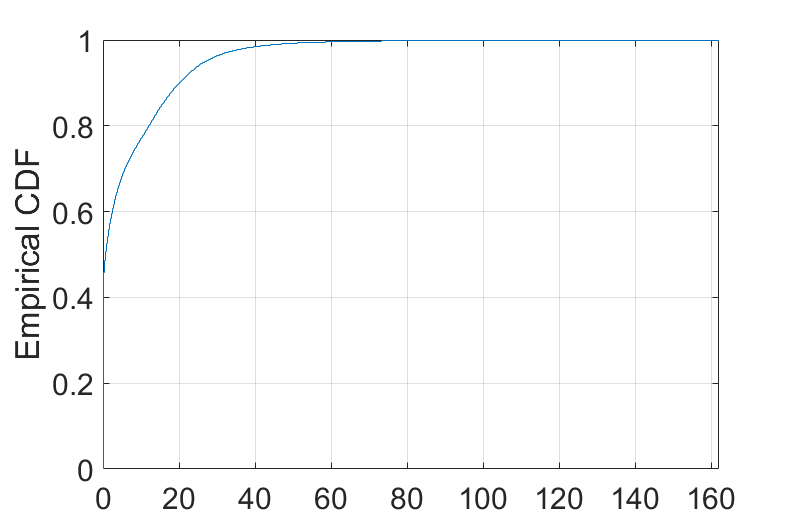}

		\vspace{-2mm}
		\caption{\small Empirical CDF of the difference between the direct and dominant path in meters.}
		\label{fig:eCDF}
	\end{figure}

		We call the straight line connecting Tx and Rx, which may go through obstacles (in our dataset, these are buildings and cars), the \emph{direct path}.
	The difference (error) between ranging with ToA measurement of the ray (range is calculated as: ToA $\times$ speed of light) and the direct path is called the \emph{NLOS (non-line-of-sight) bias}, which is by definition non-negative (zero when the link is in LOS, positive otherwise). The empirical cumulative distribution function (CDF) of the NLOS bias of the dataset is shown in Fig. \ref{fig:eCDF}. We see that $47\%$ of the links in the \emph{RadioToASeer} are in LOS.
 
	We argue that even though the dominant path may be a NLOS path, and hence introduces an NLOS bias, it is quasi-optimal to use the ToA of such paths in ToA ranging-based algorithms. First, if the dominant path is in LOS, then it is by definition the shortest path and the bias is zero. Second, in the NLOS situation, the NLOS bias of a dominant path is lower than that of a potential free space path between the same Tx and Rx, that undergoes reflections to reach its destination. Thus, relying on ToAs of the dominant paths would yield better performances than the ToAs of the other free space paths.\looseness=-1
	
	Notice that in NLOS conditions using the estimated range (length) of the direct path  would also incur several problems, due to the penetration of the obstacles. In an urban scenario, the associated direct path of an NLOS link is usually subject to blockage by numerous buildings and empirical evidence shows that loss due to penetration through a building is around 15-20 dB  \cite{585GHzRappaport}. Hence, the received power of an NLOS direct path may go easily below the detection threshold for devices with regular noise figures and SNR requirements. 

    Furthermore, as shown in \cite{LocUNetTWC}, the resolvability of such a direct path is very unlikely. So, it is reasonable to use the dominant paths instead of the direct paths. 
		
	 Thus, overall, evaluating the ToA ranging based-methods on this dataset yields essentially a best case of what is possible with ToA ranging in an urban environment.  

\end{appendices}
\bibliographystyle{IEEEtran}
 
 \bibliography{pub}

\vspace{-8mm}
\begin{IEEEbiographynophoto}{\c{C}a\u{g}kan Yapar} [S] (cagkan.yapar@campus.tu-berlin.de) received the B.Sc. degree in electrical and electronics engineering from Boğaziçi University, in 2012, and the M.Sc. degree in electrical engineering and information technology from TU München, in 2015. He is currently a Doctoral Researcher with the Communications and Information Theory Chair, TU Berlin, Germany. His research interests include communications, machine learning, information theory and signal processing.
\end{IEEEbiographynophoto}
\vspace{-4mm}
\begin{IEEEbiographynophoto}{Ron Levie} (levieron@technion.ac.il) received the Ph.D. degree in mathematics in 2018, from Tel Aviv University, Israel. During 2018-2020, he was a postdoctoral researcher with the Research Group Applied Functional Analysis, Institute of Mathematics, TU Berlin, Germany. During 2021-2022 he was a postdoctoral researcher in the Bavarian AI Chair for Mathematical Foundations of Artificial Intelligence, Department of Mathematics, LMU Munich, Germany. Since 2022, he is an assistant professor (senior lecturer) at the Faculty of Mathematics, Technion - Israel Institute of Technology.
His current research interests are in theory of deep learning, geometric deep learning, explainability of deep learning, signal processing, and applied harmonic analysis.
\end{IEEEbiographynophoto}
\vspace{-4mm}
\begin{IEEEbiographynophoto}{Gitta Kutyniok} [SM] (kutyniok@math.lmu.de) 
earned in 1996 a diploma in mathematics and computer science at Paderborn University. She completed her doctorate (Dr. rer. nat.) at Paderborn in 2000. Her dissertation, Time-Frequency Analysis on Locally Compact Groups, was supervised by Eberhard Kaniuth.
From 2000 to 2008 she held short-term positions at Paderborn University, the Georgia Institute of Technology, the University of Giessen, Washington University in St. Louis, Princeton University, Stanford University, and Yale University. In 2006 she earned her habilitation in Giessen, in 2008 she became a full professor at Osnabrück University, and in 2011 she was given the Einstein Chair at the Technical University of Berlin. In 2018 she added courtesy affiliations with computer science and electrical engineering at TU Berlin and an adjunct faculty position at the University of Tromsø. In October 2020 she moved to the Ludwig-Maximilians-Universität München, where she holds a Bavarian AI Chair.
\end{IEEEbiographynophoto}
\vspace{-4mm}
\begin{IEEEbiographynophoto}{Giuseppe Caire} [F] (caire@tu-berlin.de) is currently an Alexander von Humboldt Professor with the Faculty of Electrical Engineering and Computer Science at the Technical University of Berlin, Germany. He received the Jack Neubauer Best System Paper Award from the IEEE Vehicular Technology Society in 2003; the IEEE Communications Society and Information Theory Society Joint Paper Award in 2004 and in 2011; the Okawa Research Award in 2006; the Alexander von Humboldt Professorship in 2014; the Vodafone Innovation Prize in 2015; an ERC Advanced Grant in 2018; the Leonard G. Abraham Prize for best IEEE JSAC paper in 2019; and the IEEE Communications Society Edwin Howard Armstrong Achievement Award in 2020. He is a recipient of the 2021 Leibniz Prize of the German National Science Foundation (DFG). He has been a Fellow of IEEE since 2005. He was President of the IEEE Information Theory Society in 2011.
\end{IEEEbiographynophoto}

\vfill

\end{document}